\documentclass[reprint,amsmath,amssymb,aps,prb,showpacs,floatfix,superscriptaddress]{revtex4-1}

\usepackage{graphicx}
\usepackage{bm}
\usepackage{multirow}
\usepackage{verbatim}
\usepackage{longtable}
\usepackage{booktabs}
\usepackage{color}
\usepackage{nccmath}
\usepackage{siunitx} 
\usepackage{hyperref}
\usepackage{colortbl}
\AtBeginDocument{
\heavyrulewidth=.08em
\lightrulewidth=.05em
\cmidrulewidth=.03em
\belowrulesep=.65ex
\belowbottomsep=0pt
\aboverulesep=.4ex
\abovetopsep=0pt
\cmidrulesep=\doublerulesep
\cmidrulekern=.5em
\defaultaddspace=.5em
}

\renewcommand{\figurename}{Fig.}

\newcommand\abs[1]{\lvert#1\rvert}

\usepackage[usenames,dvipsnames]{xcolor}

\begin{document}

\title{Indistinguishable and efficient single photons from a quantum dot in a planar nanobeam waveguide}

\author{Gabija Kir{\v{s}}ansk{\.{e}}}
\affiliation{Niels Bohr Institute, University of Copenhagen, Blegdamsvej 17, DK-2100 Copenhagen, Denmark}
\author{Henri Thyrrestrup}
\affiliation{Niels Bohr Institute, University of Copenhagen, Blegdamsvej 17, DK-2100 Copenhagen, Denmark}
\author{Rapha\"{e}l S.~Daveau}
\affiliation{Niels Bohr Institute, University of Copenhagen, Blegdamsvej 17, DK-2100 Copenhagen, Denmark}
\author{Chris L.~Dree{\ss}en}
\affiliation{Niels Bohr Institute, University of Copenhagen, Blegdamsvej 17, DK-2100 Copenhagen, Denmark}
\author{Tommaso Pregnolato}
\affiliation{Niels Bohr Institute, University of Copenhagen, Blegdamsvej 17, DK-2100 Copenhagen, Denmark}
\author{Leonardo Midolo}
\affiliation{Niels Bohr Institute, University of Copenhagen, Blegdamsvej 17, DK-2100 Copenhagen, Denmark}
\author{Petru Tighineanu}
\affiliation{Niels Bohr Institute, University of Copenhagen, Blegdamsvej 17, DK-2100 Copenhagen, Denmark}
\author{Alisa Javadi}
\affiliation{Niels Bohr Institute, University of Copenhagen, Blegdamsvej 17, DK-2100 Copenhagen, Denmark}
\author{S{\o}ren Stobbe}
\affiliation{Niels Bohr Institute, University of Copenhagen, Blegdamsvej 17, DK-2100 Copenhagen, Denmark}
\author{R{\"u}diger Schott}
\affiliation{Lehrstuhl f{\"u}r Angewandte Festk{\"o}rperphysik, Ruhr-Universit{\"a}t Bochum, Universit{\"a}tsstrasse 150, D-44780 Bochum, Germany}
\author{Arne Ludwig}
\affiliation{Lehrstuhl f{\"u}r Angewandte Festk{\"o}rperphysik, Ruhr-Universit{\"a}t Bochum, Universit{\"a}tsstrasse 150, D-44780 Bochum, Germany}
\author{Andreas D.~Wieck}
\affiliation{Lehrstuhl f{\"u}r Angewandte Festk{\"o}rperphysik, Ruhr-Universit{\"a}t Bochum, Universit{\"a}tsstrasse 150, D-44780 Bochum, Germany}
\author{Suk In Park}
\affiliation{Center for Opto-Electronic Convergence Systems, Korea Institute of Science and Technology, Seoul 136-791, Korea}
\author{Jin D.~Song}
\affiliation{Center for Opto-Electronic Convergence Systems, Korea Institute of Science and Technology, Seoul 136-791, Korea}
\author{Andreas V.~Kuhlmann}
\affiliation{Department of Physics, University of Basel, Klingelbergstrasse 82, CH-4056 Basel, Switzerland}
\author{Immo S\"{o}llner}
\affiliation{Department of Physics, University of Basel, Klingelbergstrasse 82, CH-4056 Basel, Switzerland}
\author{Matthias C. L\"{o}bl}
\affiliation{Department of Physics, University of Basel, Klingelbergstrasse 82, CH-4056 Basel, Switzerland}
\author{Richard J.~Warburton}
\affiliation{Department of Physics, University of Basel, Klingelbergstrasse 82, CH-4056 Basel, Switzerland}
\author{Peter Lodahl}
\affiliation{Niels Bohr Institute, University of Copenhagen, Blegdamsvej 17, DK-2100 Copenhagen, Denmark}

\begin{abstract}
We demonstrate a high-purity source of indistinguishable single photons using a quantum dot embedded in a nanophotonic waveguide. The source features a near-unity internal coupling efficiency and the collected photons are efficiently coupled off-chip by implementing a taper that adiabatically couples the photons to an optical fiber. By quasi-resonant excitation of the quantum dot, we measure a single-photon purity larger than \SI{99.4}{\percent} and a photon indistinguishability of up to \SI{94\pm 1}{\percent} by using \emph{p}-shell excitation combined with spectral filtering to reduce photon jitter. A temperature-dependent study allows pinpointing the residual decoherence processes notably the effect of phonon broadening. Strict resonant excitation is implemented as well as another mean of suppressing photon jitter, and the additional complexity of suppressing the excitation laser source is addressed.  The study opens a clear pathway towards the long-standing goal of a fully deterministic source of indistinguishable photons, which is integrated on a planar photonic chip.
\end{abstract}

\maketitle 


\section{Introduction}

\begin{figure}
\begin{center}
\includegraphics[width=0.85\columnwidth]{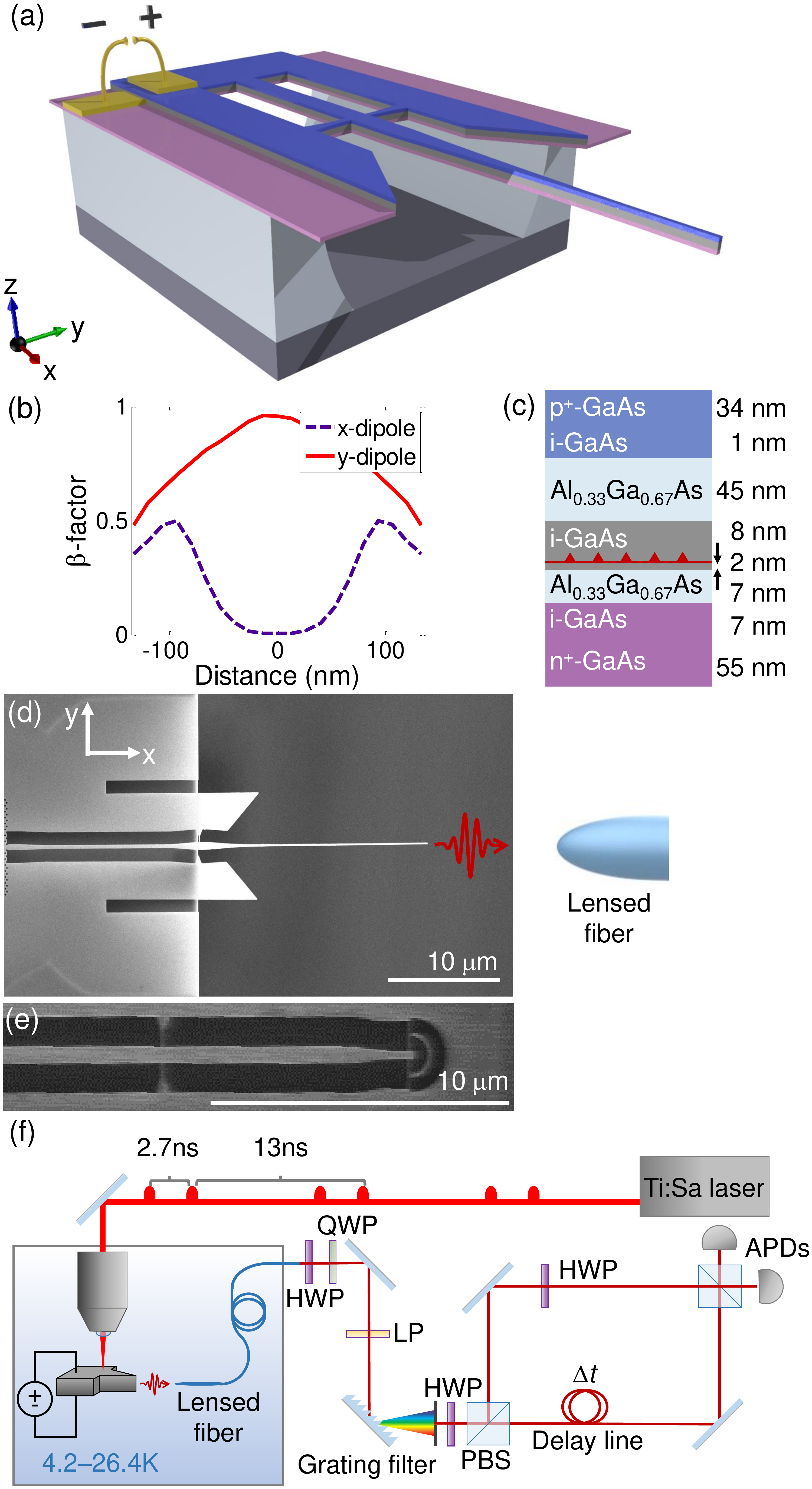}
\caption{Two-photon interference experiment on a QD embedded in a tapered nanobeam waveguide. (a) Sketch of the device. (b) Calculated  $\beta$-factor for the two dipole orientations along the x- and y-axes of the nanobeam waveguide as a function of distance from the waveguide center along the y-axis. (c) Layer structure of the wafer used for sample fabrication, where the intrinsic layer in the middle of a \emph{p-i-n} diode contains InGaAs QDs. (d) Scanning-electron micrograph of the waveguide sample with adiabatically tapered outcouplers. The emitted photons are collected by a lensed single-mode fiber. (e) Scanning-electron micrograph of the waveguide sample terminated by a circular grating outcoupler guiding the emitted photons out of plane. (f) Schematics of the experimental setup to measure the indistinguishability of single photons consecutively emitted with variable time delay $\Delta{t}$. The photoluminescence of the QD is tuned via the applied electric field. The emitted photons are collected with a lensed fiber and sent to a HOM interferometer for correlation measurements. (HWP: half-wave plate, QWP: quarter-wave plate, LP: linear polarizer, PBS: polarizing beam splitter APDs: avalanche photodiodes).}
\label{fig1}
\end{center}
\end{figure}

A truly on-demand source of coherent single photons  is the essential quantum hardware behind many photonic quantum-information applications including device-independent quantum cryptography~\cite{Diamanti2016,Sangouard2012} and quantum simulations~\cite{Aspuru-Guzik2012}, or more daring, a full-scale photonic quantum computer~\cite{Kok2007,Rudolph2016} or a photonic quantum network~\cite{Kimble2008,Lodahl2017}.
The demands on source performance depend on the actual application in mind, and currently the first proof-of-concept demonstrations emerge. At present the state-of-the-art in the field is the achievement of 10-photon entanglement~\cite{Wang2016a} and quantum simulations with 3--4 photons~\cite{Spring2013,Broome2013}. A highly promising  route of extending beyond this performance applies self-assembled quantum dots (QDs) as single-photon emitters~\cite{Michler2000}  embedded in photonic nanostructures to enhance light-matter interaction \cite{Lodahl2015}. This platform has matured significantly in recent years~\cite{Santori2002,Laurent2005,Lund-Hansen2008,Claudon2010,Prechtel2013,Kuhlmann2013,Arcari2014,Kuhlmann2015,Lodahl2015,Gschrey2015,Somaschi2016,Loredo2016,Wang2016,Ding2016,Thoma2016}. So far, much progress has been obtained on micropillar cavities and nanowires where the collected photons are coupled vertically out of the structure in a confocal microscopy setup. Planar nanophotonic waveguides offer the opportunity of increasing the single-photon coupling efficiency to near unity~\cite{Arcari2014}. Importantly, the waveguide-integrated platform provides a route to on-chip photonic quantum-information processing. It remains to be demonstrated that highly coherent single photons can be generated on this platform, where the presence of surfaces near the QD may lead to decoherence~\cite{Houel2012}. Indeed the thin ($\sim$~\SI{160}{\nano\meter}) and narrow ($\sim$~\SI{300}{\nano\meter}) waveguide structures imply that the embedded QDs are unavoidably close to doped semiconductor material and semiconductor-air interfaces.

We present the demonstration of a highly coherent single-photon source based on an electrically controlled QD integrated in a high-efficiency nanophotonic waveguide. Pulsed quasi-resonant and strict resonant excitation of the QD in the waveguide is applied to deterministically operate the single-photon source, whereby the indistinguishability of the emitted photons can be directly determined as opposed to a continuous-wave excitation experiment~\cite{Kalliakos2014,Kalliakos2016}. Single-photon purity exceeding \SI{99.4}{\percent} is demonstrated together with a two-photon interference visibility of \SI{62}{\percent} for quasi-resonant excitation. These values are limited primarily by photon jitter associated with the relaxation from the excited \emph{p}-shell and into the ground state. On a different sample, we show that the spectral filtering allows suppressing photon jitter significantly and observe an indistinguishability as high as  \SI{94\pm 1}{\percent}. Strict resonant excitation is implemented as an alternative method of suppressing photon jitter without the need to spectrally filter the output, while instead the efficient suppression of the pump light is required.  Finally, it is demonstrated how the emitted photons can efficiently be coupled off chip to an optical fiber by the implementation of a tapered waveguide section. Our work paves the way for a fully deterministic source of indistinguishable photons for scalable quantum-information processing applications.

The paper is organized as follows. In Sec.~II the structure design, the experimental methods, and the results of the quasi-resonant indistinguishability measurements are presented. In Sec.~III we discuss the outcoupling taper design and the results of the source efficiency characterization performed. In Sec.~IV we present the Hong-Ou-Mandel (HOM) experiment results on a strictly resonantly excited QD embedded in a nanobeam waveguide terminated by a grating outcoupler. We present the conclusions in Sec.~V. In Appendix~A we discuss the fitting routine allowing for reliable extraction of photon indistinguishability from a pulsed HOM experiment. Details on the analysis of the resonance fluorescence HOM data with continuous-wave (CW) background are included in Appendix~B.

\section{Photon indistinguishability under quasi-resonant excitation}

\subsection{Structure design: Device~A}

The investigated device (Device~A) consists of a suspended waveguide with integrated metal contacts in order to apply an external bias across the InGaAs QD to tune and stabilize the transition. The 300-nm-wide suspended waveguides are fabricated on a \emph{p-i-n} GaAs wafer (cf.~\figurename~\ref{fig1}(c) for the layer structure) using electron-beam lithography followed by dry and wet etching processes, following the methods described in Ref.~\onlinecite{Midolo2015}. Figures~\ref{fig1}(d) shows a scanning-electron microscope image of the samples terminated with a taper, e.g., Device~A, where a  precise cleaving method is implemented to obtain tapers protruding from the edge of the sample such that the optical mode in the taper can freely expand without reflecting from the device substrate. The internal efficiency of the source is quantified by the $\beta$-factor, which is the single-photon coupling efficiency into the waveguide mode~\cite{Lodahl2015}. Figure~\ref{fig1}(b) displays the spatial dependance of the simulated $\beta$-factor in the waveguide for the two relevant dipole orientations. At the center of the waveguide the optical transition polarized along the y-dipole couples to the main waveguide mode with \SI{96}{\percent} efficiency, whereas the orthogonal dipole is polarization-filtered in the structure. We implement a way to out-couple photons from a waveguide on a photonic chip directly to a single-mode fiber, thus discarding the need for a confocal setup. Further details on the taper design are discussed in Sec.~III.

\subsection{Experimental conditions}

For the optical measurements under quasi-resonant excitation, Device~A is mounted on a three-axis piezoelectric stage in a liquid-helium bath cryostat and cooled to temperatures between 4.2--\SI{26.4}{\kelvin} (see Fig.~\ref{fig1}(f)). The QD is excited with a picosecond-pulsed Ti:Sapphire laser at a repetition rate of \SI{76}{\mega\hertz} and focused through a microscope objective with NA~$= 0.55$. The laser polarization is set to the orientation resulting in the highest count rate of the emitted photons. A delay is introduced in the laser path to allow for a periodic excitation with double pulses delivered with a time separation of \SI{2.7}{\nano\second}. The emitted photons are collected by a lensed single-mode fiber mounted on a separate stack of piezoelectric stages allowing for the taper-fiber spatial alignment and guided to a detection setup. Polarization scrambling in the single-mode fibers is corrected by the polarization optics after the lensed fiber. A grating filtering setup is implemented (spectral resolution: \SI{70}{\pico\meter}/\SI{25}{\giga\hertz}, transmission throughput: \SI{27}{\percent}) in order to remove the phonon sidebands of the single photons, whereby only the spectrally narrow zero-phonon line is analyzed. The emitted photons are characterized in a HOM interferometer~\cite{Hong1987} illustrated in~\figurename~\ref{fig1}(e). Here the spectrally filtered photons are sent to an asymmetric Mach-Zehnder interferometer designed to record the quantum interference of subsequently emitted photons. The photon indistinguishability is obtained from the two-photon auto-correlation function constructed by correlating two single-photon detectors.

\subsection{Single-photon purity}

\begin{figure}
\begin{center}
\includegraphics[width=0.9\columnwidth]{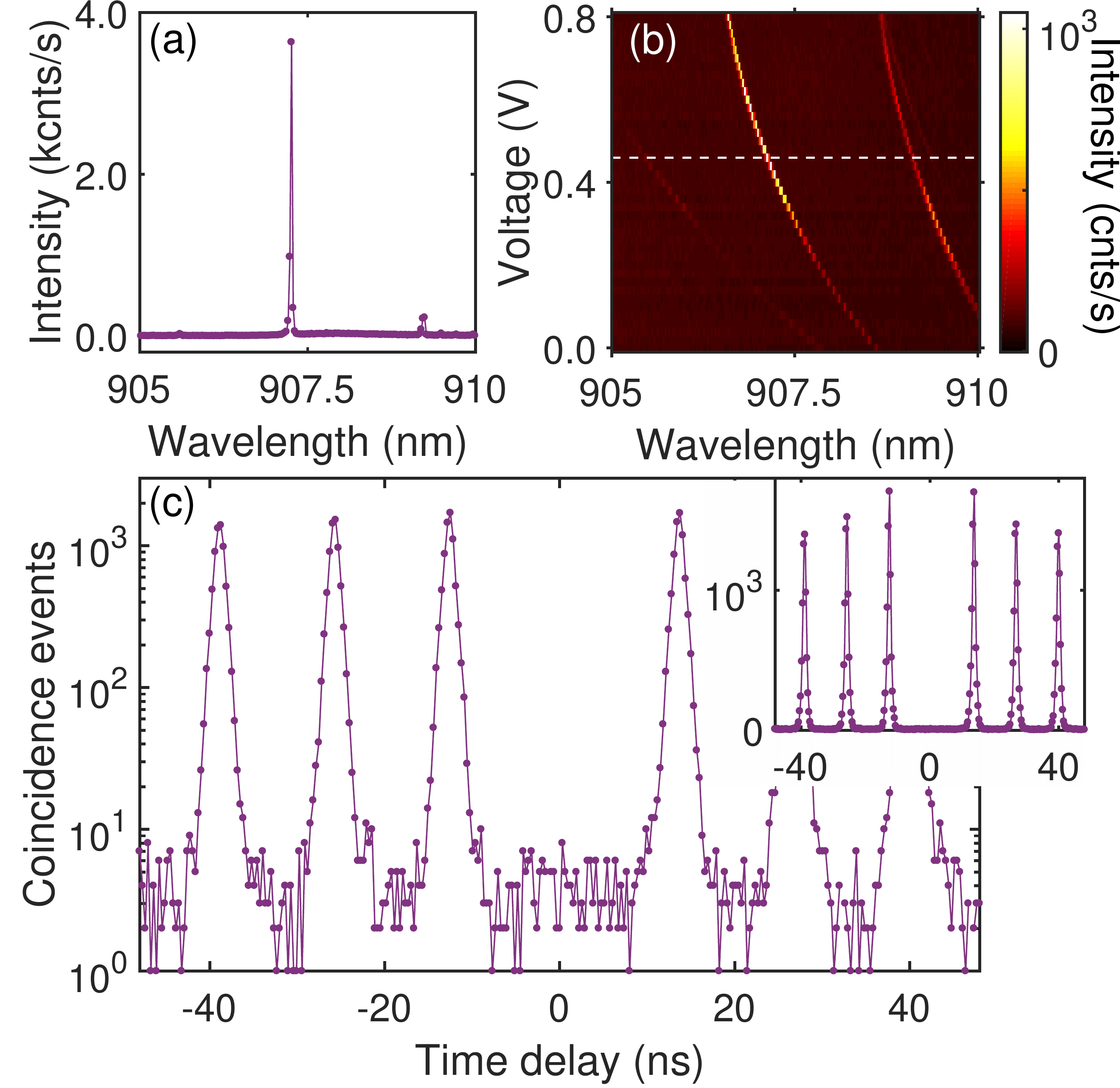}
\caption{Single-photon emission from an electrically controlled QD in Device~A under pulsed \emph{p}-shell excitation at \SI{4.2}{\kelvin}. (a) Photoluminescence spectrum of the QD emitting at \SI{907.4}{\nano\meter} at \SI{0.46}{\volt} bias (indicated by the white dashed line in (b)). The width of the emission line is limited by the resolution of the spectrometer. (b) Photoluminescence map of the QD versus the applied voltage measured at an excitation level of 80 \% of the saturation power. An extra line from the same QD is identified at a lower voltage and the pair most likely is the $X^0$ and $X^{+}$ transitions. The  circularly polarized $X^{+}$ couples less efficiently to the nanobeam waveguide resulting in a weaker optical signature than $X^0$. (c) Intensity-correlation histogram from the QD under \emph{p}-shell excitation on a logarithmic scale. Nearly ideal single-photon emission is demonstrated by the vanishing multi-photon probability at zero time delay $g^2(0) < \num{0.006}$. The inset shows the data in (c) on a linear scale.}
\label{fig3}
\end{center}
\end{figure}

\begin{figure}[bp]
\begin{center}
\includegraphics[width=0.85\columnwidth]{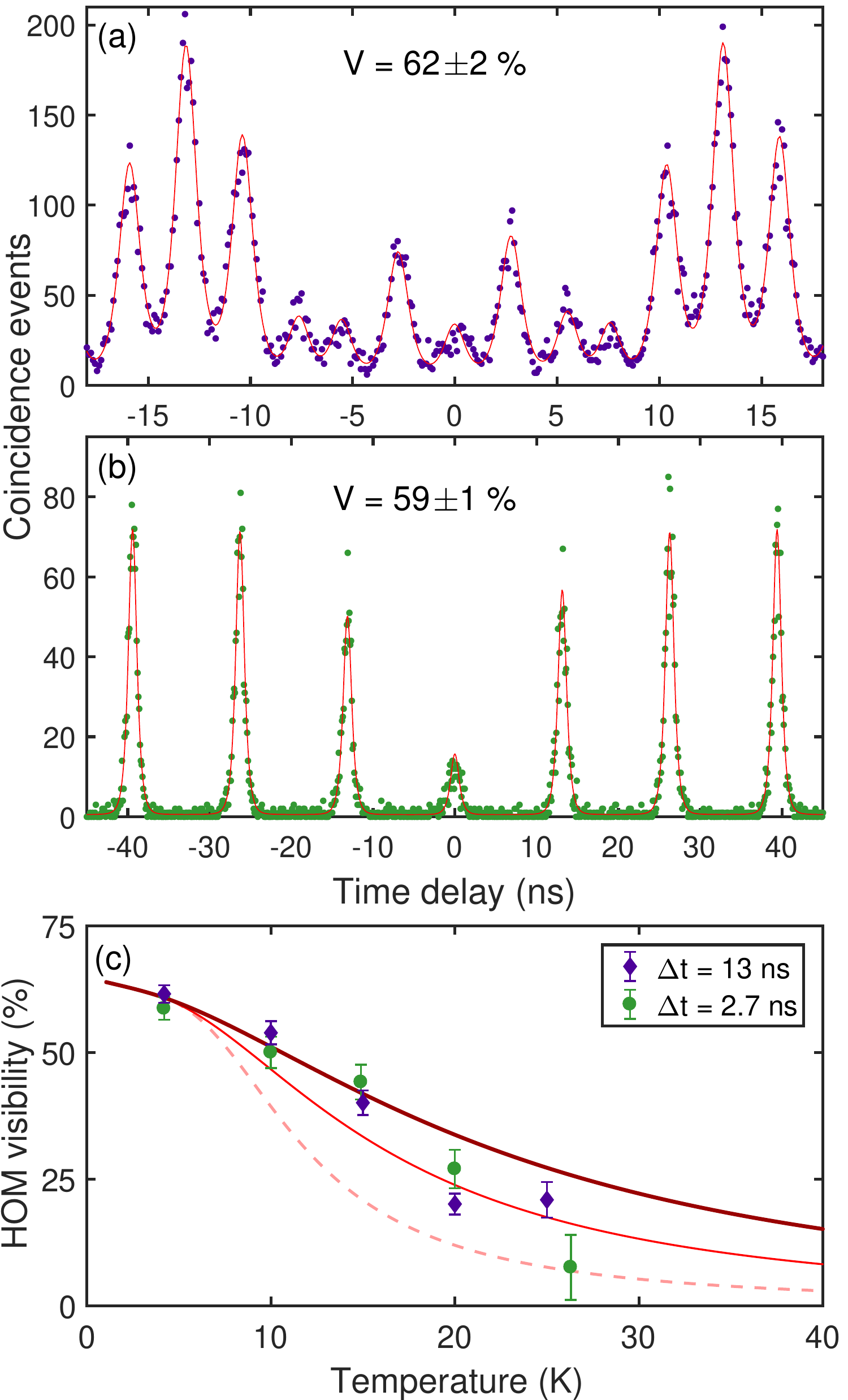}
\caption{Recorded correlation histograms in the HOM interferometer of Device~A for a successive time separation of emitted photons of \SI{2.7}{\nano\second} (a) and \SI{13}{\nano\second} (b), respectively, measured at
\SI{4.2}{\kelvin} and at $0.8P_{\mathrm{sat}}$ excitation power. The data points present the raw data without any background correction after integration over \SI{2.5}{\hour} and \SI{1}{\hour}, respectively. We note that the outcoupling and detection efficiency was not optimized in this experiment. The red line displays the modeling of the data. The visibility $V$ is extracted by fitting the data with the convolution of a single-exponential decay (QD decay) and a Voigt function (photodetector response), see Appendix~A. (c) Measured temperature dependence of the visibility for the two time separations. The error bars are calculated by error propagation of the fitted parameter errors.  The curves are obtained from the theory presented in Ref.~\onlinecite{Tighineanu2016}, and for spherical QDs of radii 2, 3, and \SI{4}{\nano\meter}~\cite{Fry2000,Bruls2003} and according hole (electron) energy splitting between \emph{s}- and \emph{p}-shell of 15, 20, and \SI{30}{\milli\electronvolt} (30, 40, and \SI{60}{\milli\electronvolt}) ~\cite{Raymond2004,Blokland2007} resulting in the dashed, thin solid, and thick solid lines, respectively.
 }
\label{fig4}
\end{center}
\end{figure}

The purity of single-photon emission can be characterized in a Hanbury-Brown Twiss measurement of the photon autocorrelation function. The representative experimental data reported here were recorded on Device~A. \figurename~\ref{fig3}(a) shows a typical emission spectrum for a QD quasi-resonantly excited at a wavelength of \SI{888}{\nano\meter}, which may correspond to either \emph{p}-shell or LO-phonon assisted excitation. A constant bias voltage of \SI{0.46}{\volt} is applied to the QD in order to achieve the highest photon count rate, as deduced from the photoluminescence-voltage map, shown in \figurename~\ref{fig3}(b). The measured autocorrelation function of $g^2(0) < \num{0.006}$ corresponds to an excellent single-photon purity larger than \SI{99.4}{\percent}, cf.~\figurename~\ref{fig3}(c), where $g^2(0)$ is obtained as the ratio of integrated counts of the fitted peak centered at zero time delay relative to the peak area in the uncorrelated limit of a long time delay. This accounts for the weak blinking of emission (at a preparation efficiency of \SI{75}{\percent}) found on a time scale of about \SI{40}{\nano\second}~\cite{Santori2004,Davanco2014}.

\subsection{Two-photon interference}

The indistinguishability of the single photons can be tested by two-photon interference measurements of consecutively emitted photons. Figure~\ref{fig4}(a) shows a correlation histogram of the two-photon coincidence events acquired on Device A. The five-peak structure of the histogram stems from the arrival of the photons at different time intervals. The central peak corresponds to the situation where two photons meet up on the beamsplitter at the same time and would vanish for completely indistinguishable photons. The degree of indistinguishability is quantified from the visibility $(V)$ in the HOM experiment, which can be extracted from the relative area of the central peak compared to neighboring peaks~\cite{Santori2002,Loredo2016}.
We implement a rigorous fitting routine, which takes into account the exponential decay of the emitter, the measured instrument response function, and Poissonian counting statistics. This is essential in order to reliably extract $V$, and is not generally performed in the literature~\cite{He2013,Gazzano2013,Madsen2014,Gschrey2015,Somaschi2016,Ding2016,Thoma2016,Grange2017}, with the notable exception of Ref.~\onlinecite{Kambs2016}. In Appendix~A the details of the data analysis are presented, and it is found that an overestimation of $V$ of up to \SI{30}{\percent} may be done if the data analysis often presented in the literature is implemented. Figure~\ref{fig4}(a) displays the data for a time delay in between photons of \SI{2.7}{\nano\second} where we obtain $V=\SI{62\pm 2}{\percent}$. The experiment was also repeated for a time separation  of \SI{13}{\nano\second}, cf.~\figurename~\ref{fig4}(b), where we extract $V=\SI{59\pm 1}{\percent}$, which agrees within the error-bars of the measurement with the \SI{2.7}{\nano\second} data.

The residual decoherence processes found in the measurements can be attributed to two different processes: time-jitter induced by the relaxation of the carrier from the quasi-resonant excitation to the QD ground state expressed by the rate $\Gamma_\mathrm{jitter}$~\cite{Kiraz2004} and temperature-dependent broadening (pure dephasing rate $\Gamma_\mathrm{ph}(T)$) of the zero-phonon line of the QD due to interaction with phonons~\cite{Muljarov2004}. The visibility can be expressed as~\cite{Kiraz2004}
\begin{equation}
V = \frac{\Gamma_\mathrm{rad}}{\left(\Gamma_\mathrm{rad} + \Gamma_\mathrm{ph}(T) \right) \left(1 + \Gamma_\mathrm{rad} / \Gamma_\mathrm{jitter}\right)},
\end{equation}
where $\Gamma_\mathrm{rad}=\SI{2.3}{\per\nano\second}$ is the measured radiative decay rate of the QD. Figure~\ref{fig4}(c) shows experimental data of the temperature dependence of the visibility. It is found to decrease significantly with temperature, which is indicative of phonon dephasing. The experimental data are compared to a theoretical model predicting the reduction of indistinguishability with temperature due to the broadening of the zero-phonon line~\cite{Tighineanu2016} for three indicative sets of parameters, cf.~\figurename~\ref{fig4}(c). We obtain $\Gamma_\mathrm{jitter}=\SI{3.7}{\nano\second}^{-1}$ similar to literature reported values~\cite{Reithmaier2014}. This contribution may be reduced by applying resonant $(\pi$-pulse) excitation or spectral filtering, as will be demonstrated in Sec. \ref{Resonant}. As an important reference point the achievable indistinguishability limitation due to phonons is $V=\SI{94}{\percent}$  at $T=\SI{4}{\kelvin}$, which is consistent with linewidth measurements \cite{Kroner2009}. Ultimately the indistinguishability will be limited by the increased broadening of the zero-phonon line due to mechanical vibrations predicted in a 1D optical system~\cite{Tighineanu2016}. This could be improved further by either cooling the sample further down or by implementing Purcell enhancement. For instance, at $T=\SI{1}{\kelvin}$ the indistinguishability would increase to \SI{98.6}{\percent} or alternatively a readily achievable Purcell factor of ten \cite{Lodahl2015} would lead to $V=\SI{98.8}{\percent}$ at $T=\SI{4}{\kelvin}$. In Sec. \ref{Resonant} we demonstrate $V=\SI{94}{\percent}.$

\section{Single-photon source efficiency}
\subsection{Efficient outcoupling taper design}

\begin{figure}
\begin{center}
\includegraphics[width=0.8\columnwidth]{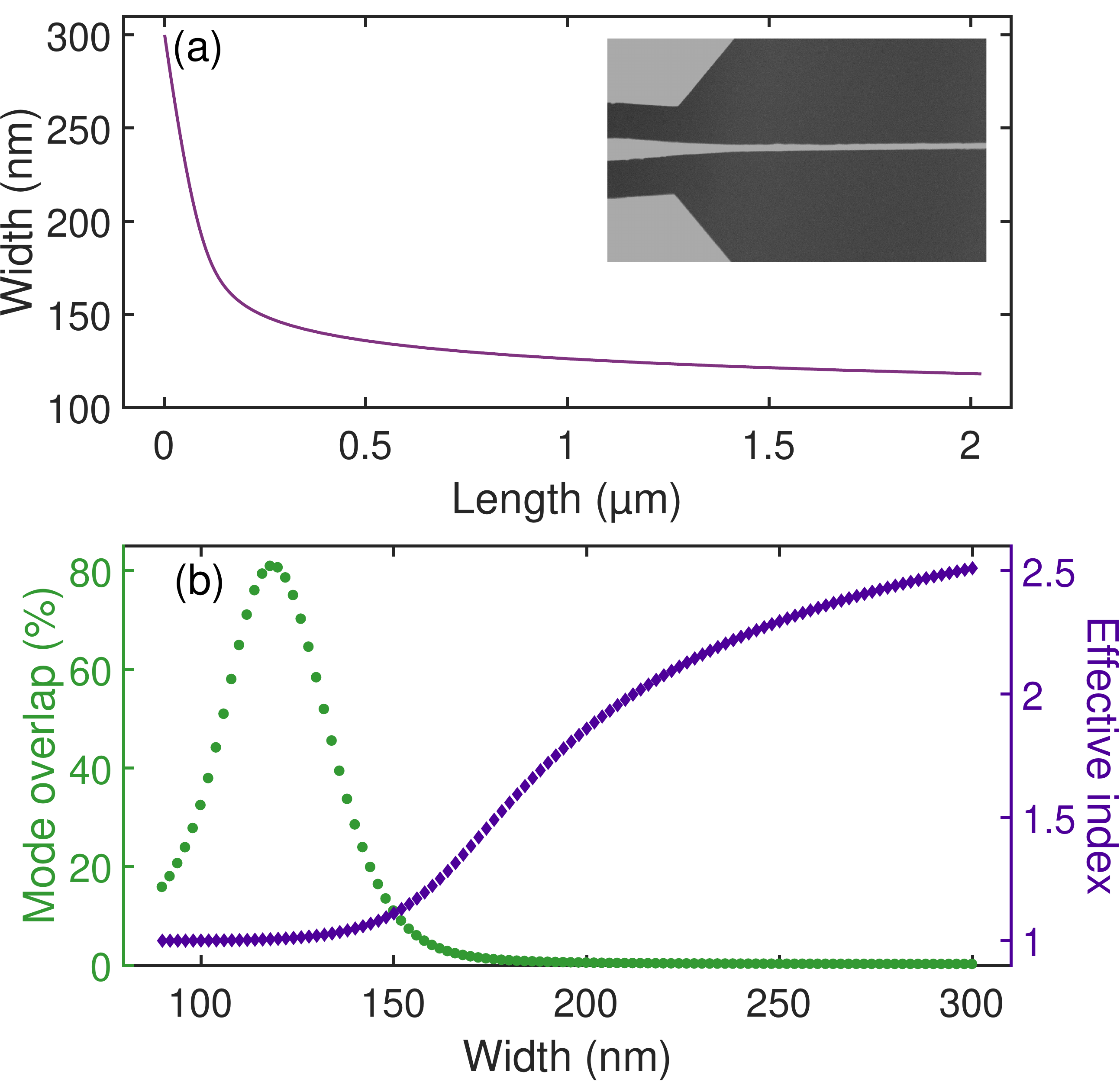}
\caption{Taper shape and properties extracted from the finite-element method simulation of a 160-nm-thick GaAs waveguide surrounded by air. (a) Taper profile (width as a function of length) obtained from the adiabatic rule with $\alpha = 1$ to emphasize the shape.  Inset---a fragment of a scanning-electron micrograph of a fabricated taper with $\alpha=10$. (b) Mode overlap between the waveguide mode and a Gaussian distribution of \SI{2.5}{\micro\meter} mode-field diameter (green dotted curve). The maximum mode overlap serves as a guideline for the optimal taper width. Effective index as a function of the waveguide width (blue points) illustrates when the optical mode is fully leaked out into air (at a width of $\lesssim \SI{150}{\nano\meter}$).}
\label{fig1SM}
\end{center}
\end{figure}

The efficient extraction of the emitted photons is implemented through an adiabatic taper outcoupler, which is presented in the following. The waveguide structures are terminated with a taper out-coupling section inspired by the work of Cohen \emph{et al.}~\cite{Cohen2013}, and adapted to GaAs for a working wavelength of \SI{940}{\nano\meter}. The efficient coupling from a sub-micrometer sized waveguide to a \SI{5}{\micro\meter} single-mode fiber core requires a redesign of the two systems to achieve good mode matching. Lensed-fibers allow to reduce the fiber mode typically to a few microns, however the dimensions of the $300 \times \SI{160}{\nano\meter}$ waveguide mode cannot be matched. This is overcome by tapering the waveguide along the propagation direction, which forces the optical mode of the waveguide to expand thereby gradually transferring the mode from GaAs with high refractive index ($n_\textrm{GaAs}$ = 3.4) to air. The waveguide is tapered nonlinearly, as shown in \figurename~\ref{fig1SM}(a), in order to achieve the required mode diameter at a working distance of a lensed fiber. This ensures that the fiber can be operated at a safe distance from the nanophotonic waveguide during experiments.

Figure \ref{fig1SM}(b) shows the effective index of the GaAs waveguide and the mode overlap as a function of waveguide width. The maximal mode overlap is achieved when the waveguide mode is almost entirely in air while still being guided, i.e., the effective index of the waveguide is close to 1. For a waveguide width of \SI{118}{\nano\meter}, which can be fabricated, the mode overlap is maximized to \SI{80}{\percent}. Efficient mode conversion in the present case from a \SI{300}{\nano\meter} to a \SI{118}{\nano\meter} waveguide, requires that the adiabatic condition is fulfilled~\cite{Snyder1984}
\begin{equation}\label{eq:adiabatic}
   \frac{\mathrm{d}w(z)}{\mathrm{d}z} \ll n_\mathrm{eff}(w) - n_\mathrm{clad},
\end{equation}
where $n_\mathrm{eff}$ ($n_\mathrm{clad}$) is the effective index of the waveguide mode (the material surrounding the waveguide). That is, the width of the waveguide along the propagation direction $w(z)$ has to change slowly to prevent coupling of the fundamental mode to higher-order or counter-propagating modes. Because of the high $n_\mathrm{eff}$ for GaAs waveguides, the effective index of the \SI{300}{\nano\meter} waveguide is relatively large compared to the refractive index of air allowing us to change the width of the waveguide relatively fast initially. We define the adiabatic factor $\alpha$ from the relation
 \begin{equation}\label{eq:adiabatic2}
 \frac{\mathrm{d}w(z)}{\mathrm{d}z} = \alpha^{-1}(n_\mathrm{eff}(w) - 1),
 \end{equation}
where $\alpha \gg 1$ is the adiabatic condition. Setting $\alpha = 1$ at first allows to find the optimal taper shape, which can then be scaled afterwards to an arbitrary length that fulfils the adiabatic criterion. Equation \eqref{eq:adiabatic2} cannot be integrated directly because of the dependence of $n_\mathrm{eff}(w)$, so instead the change in position $\Delta z_i$  between two consecutive waveguide widths $w_i$ is obtained through $\Delta z_i = \Delta w/(n_\mathrm{eff}(w_i) - 1)$. This equation defines the profile of the waveguide tapered from \SI{300}{\nano\meter} to \SI{118}{\nano\meter} as shown in \figurename~\ref{fig1SM}(a). The length of the taper is finally $\alpha \sum \Delta z_i$, which is \SI{1.8}{\micro\meter} for $\alpha = 1$. The taper fabricated on Device~A used in the quasi-resonant HOM experiments in the manuscript was designed with $\alpha = 10$. Below, we discuss the efficiency characterization results on another device (Device~B) with $\alpha = 4$.

\subsection{Efficiency characterization: Device~B}

The single-photon source efficiency has been thoroughly characterized on Device~B similar to Device~A in Fig.~\ref{fig1}(a) but with no electrical contacts. It was fabricated on an intrinsic 160-nm-thick GaAs membrane and contains several waveguide sections with a photonic-crystal mirror termination on one side and waveguide taper on the other side, making the device unidirectional. The photonic-crystal waveguide has a 5-$\mu$m-long slow-light section to enhance the light-matter coupling to reach near unity $\beta$-factor~\cite{Arcari2014}. The photonic-crystal waveguide is coupled to a 5-$\mu$m-long straight nanobeam waveguide, eventually tapered to a width of \SI{118}{\nano\meter} with an adiabatic parameter of $\alpha = 4$ according to the design introduced above.

\begin{figure}
\begin{center}
\includegraphics[width=0.9\columnwidth]{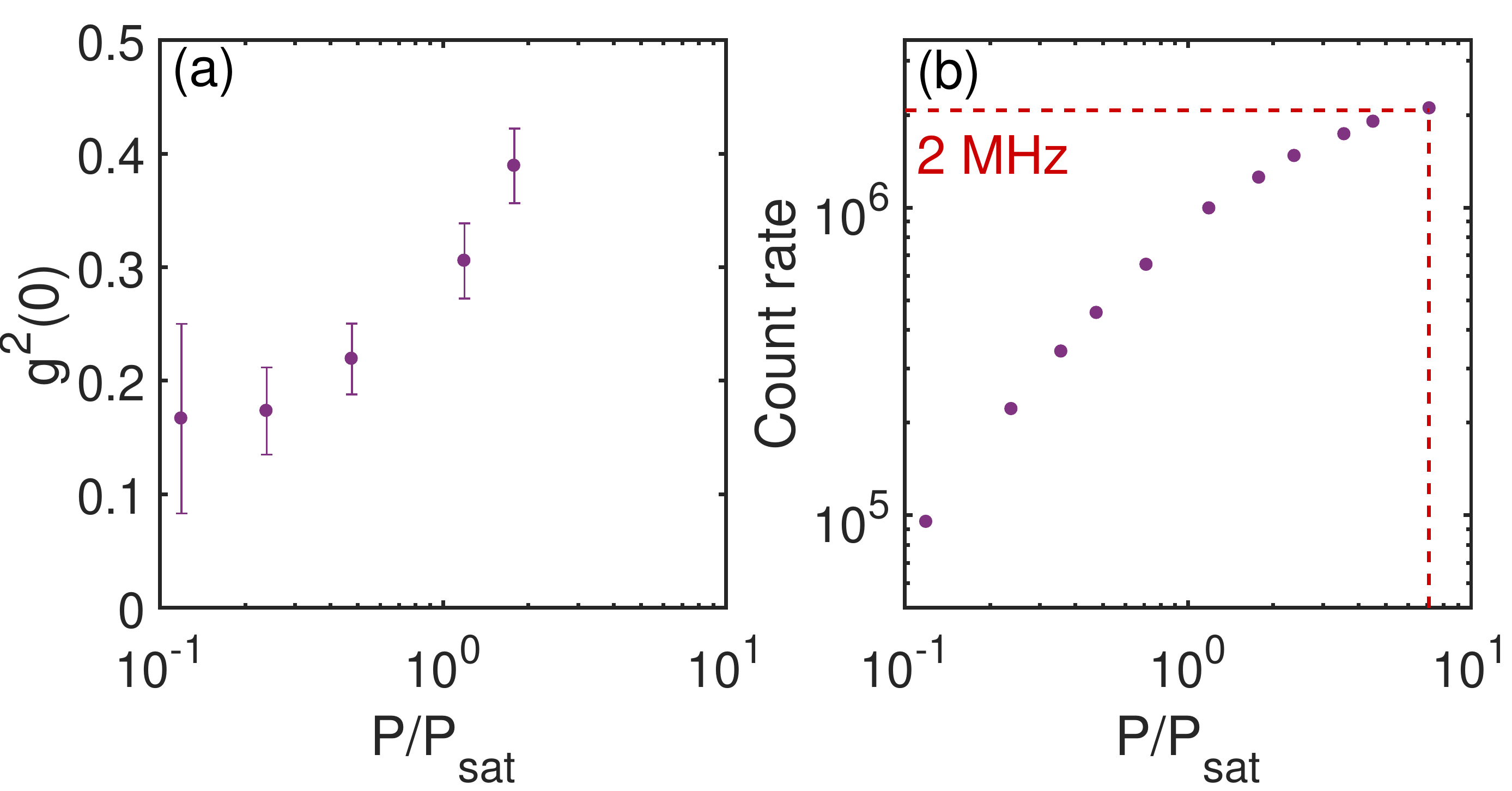}
\caption{Efficiency of a single-photon source with no electrical gates (Device B). (a) Autocorrelation function at zero time delay $g^{2}(0)$ vs excitation power measured in units of the saturation power $P_{sat}$. (b) Raw count rate on an avalanche photo-diode as a function of excitation power. A count rate of up to \SI{2}{\mega\hertz} is observed, where about half of it stems from the QD.}
\label{figEfficiency}
\end{center}
\end{figure}

The overall photon-extraction efficiency of Device~B is characterized by recording the total number of detected single-photons on an APD from a QD in the photonic crystal section. A maximum count rate of $\thicksim$\SI{2}{\mega\hertz} is obtained above saturation (cf.~Fig.~\ref{figEfficiency}). The corresponding detected single-photon efficiency near the saturation power is $\thicksim$\SI{1}{\mega\hertz} after correcting for the multi-photon probability reflected in the final value of $g^{2}(0)$ shown in Fig.~\ref{figEfficiency}. The multi-photon events stem from the emission of other QDs. This corresponds to an overall source efficiency of $\eta_\textrm{sp} = \SI{10.3}{\percent}$, where  $\eta_\textrm{sp}$ is the probability that upon excitation, the QD emits a photon that is collected by the waveguide and subsequently successfully transferred into the optical fiber. The source efficiency is obtained by accounting for all propagation losses in the fiber and detection, including APD detection efficiency (\SI{26}{\percent}), spectral filtering efficiency (\SI{30}{\percent}), and the use of two mating sleeves in between fibers ($ \SI{79}{\percent}$ each). The source efficiency is limited by a finite preparation efficiency of the QD, propagation loss in the waveguide, and the coupling efficiency off-chip. A thorough analysis of the source efficiency is presented in Ref.~\onlinecite{Daveau2016}. It may be further improved by implementing evanescent coupling from the waveguide to the fiber~\cite{Daveau2016}.

\section{Two-photon interference under resonant excitation}
\label{Resonant}
\subsection{Structure design and experimental setup: Device~C}

To experimentally test the influence of photon jitter, we have also carried out HOM measurements on QDs in a nanophotonic waveguide terminated with a grating out-coupler (Device~C, cf.~Fig.~\ref{fig1}(e)). The wafer material and the fabrication procedure of the structure and the electrical contacts is the same as for Device~A (cf.~\figurename~\ref{fig1}(c) for the layer structure). In the resonance fluorescence experiment, Device~C is mounted in a closed-cycle cryostat and cooled to 1.58-\SI{1.7}{\kelvin}, and the photons are collected from the top through an objective with NA~$=0.82$. The QD is excited by a tunable CW diode laser (Toptica CTL) polarized perpendicular to the waveguide that is tuned into resonance with the QD emission frequency. A forward bias of $\SI{0.1952}{\volt}$ is applied across the \emph{p-i-n} junction resulting in an emission frequency of $\nu=\SI{325.457}{\tera\hertz}$. The residual laser is suppressed by the spatial separation between the QD and the collection spot alone as the polarization of the output grating is co-linear with the laser polarization. A laser extinction relative to the resonant fluorescence intensity of more than 200 times is achieved under CW excitation. For the pulsed resonant HOM measurement, the QD is excited with 100-\SI{}{\pico\second}-long pulses created by electro-optical modulation of the diode laser at a rate of \SI{72.6}{\mega\hertz}. Due to a finite modulation extinction, a constant CW laser contribution is present during the measurements. The \emph{p}-shell excitation is achieved with a Ti:Sapphire laser delivering ps-long pulses at a repetition rate of \SI{76}{\mega\hertz}.  The HOM interferometer used for the measurements with both excitation schemes is similar to the one sketched in Fig.~\ref{fig1}(f), now with a time delay of $\Delta t=\SI{13.8}{\nano\second}$ and $R=0.5$ and $T=0.5$. In addition to the \SI{70}{\pico\meter}/\SI{25}{\giga\hertz} grating filter (\SI{72}{\percent} transmission) the photons are tightly filtered by an etalon with a bandwidth of \SI{3}{\giga\hertz} (\SI{90}{\percent} transmission).

\subsection{Quasi-resonant excitation and spectral filtering}

\begin{figure}
\begin{center}
\includegraphics[width=0.9\columnwidth]{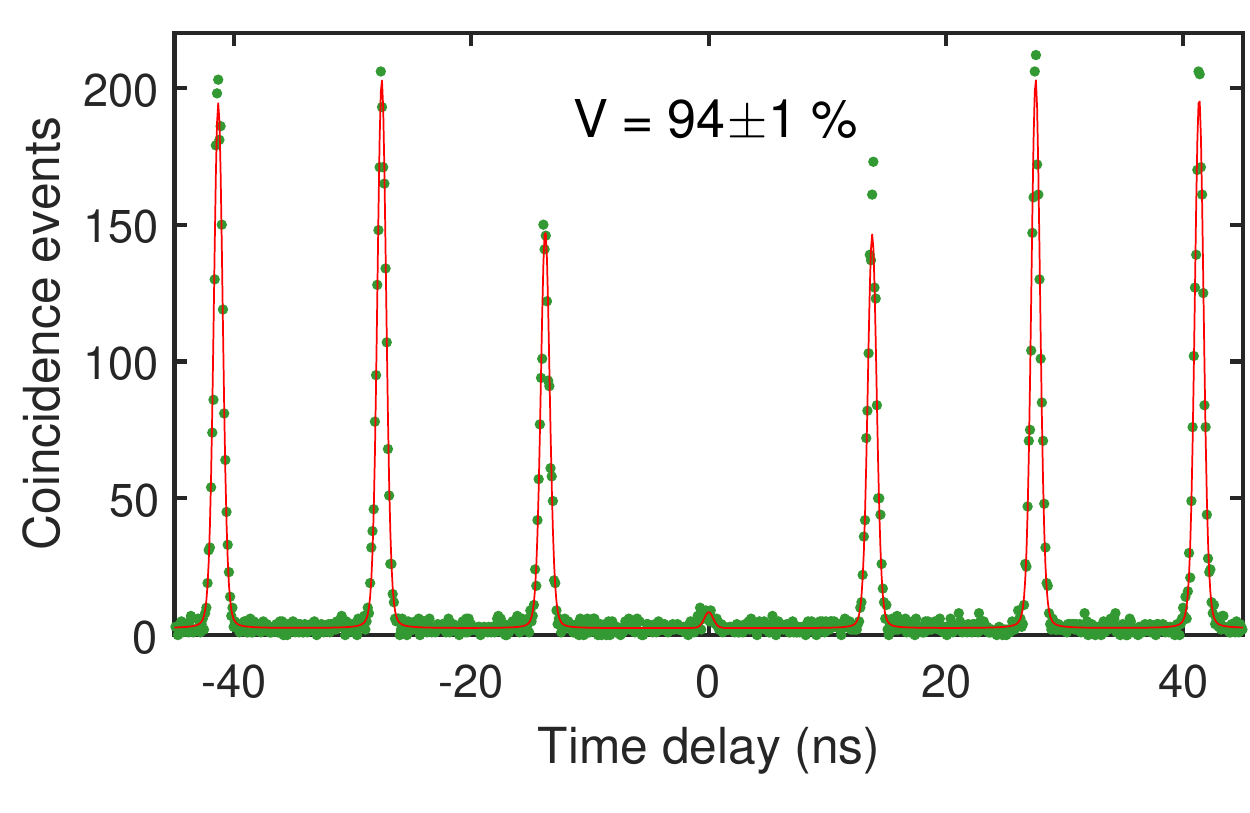}
\caption{Coincidences histogram of the HOM interference data recorded on Device~C under quasi-resonant excitation for a successive time separation of emitted photons of $\sim \SI{13}{\nano\second}$, measured at
\SI{1.7}{\kelvin} and at $P_{\mathrm{sat}}$. The data points present the raw data without any background correction. The fit of the experimental data to theory is shown by the red line.
 }
\label{figpshell}
\end{center}
\end{figure}

The implementation of spectral filtering offers one approach to improve indistinguishability. Figure~\ref{figpshell} shows a pulsed HOM interference measurement for the QD in Device~C recorded under \emph{p}-shell excitation  and by implementing a \SI{3}{\giga\hertz} bandpass etalon filter. We observe a degree of indistinguishability as high as \SI{94\pm 1}{\percent}, which approaches the fundamental limit set by the phonon broadening in nanobeam waveguides ~\cite{Tighineanu2016}. This value could potentially be improved even further by implementing rate enhancement, either through the Purcell effect or a large oscillator strength. For comparison, we observe an indistinguishability of \SI{77}{\percent} for this QD in the absence of spectral filtering, where the observed reduction could be partly due to the minor neighboring peak observed in Fig. \ref{figLinewidth}.

\subsection{Resonant excitation}

\begin{figure}
\begin{center}
\includegraphics[width=0.9\columnwidth]{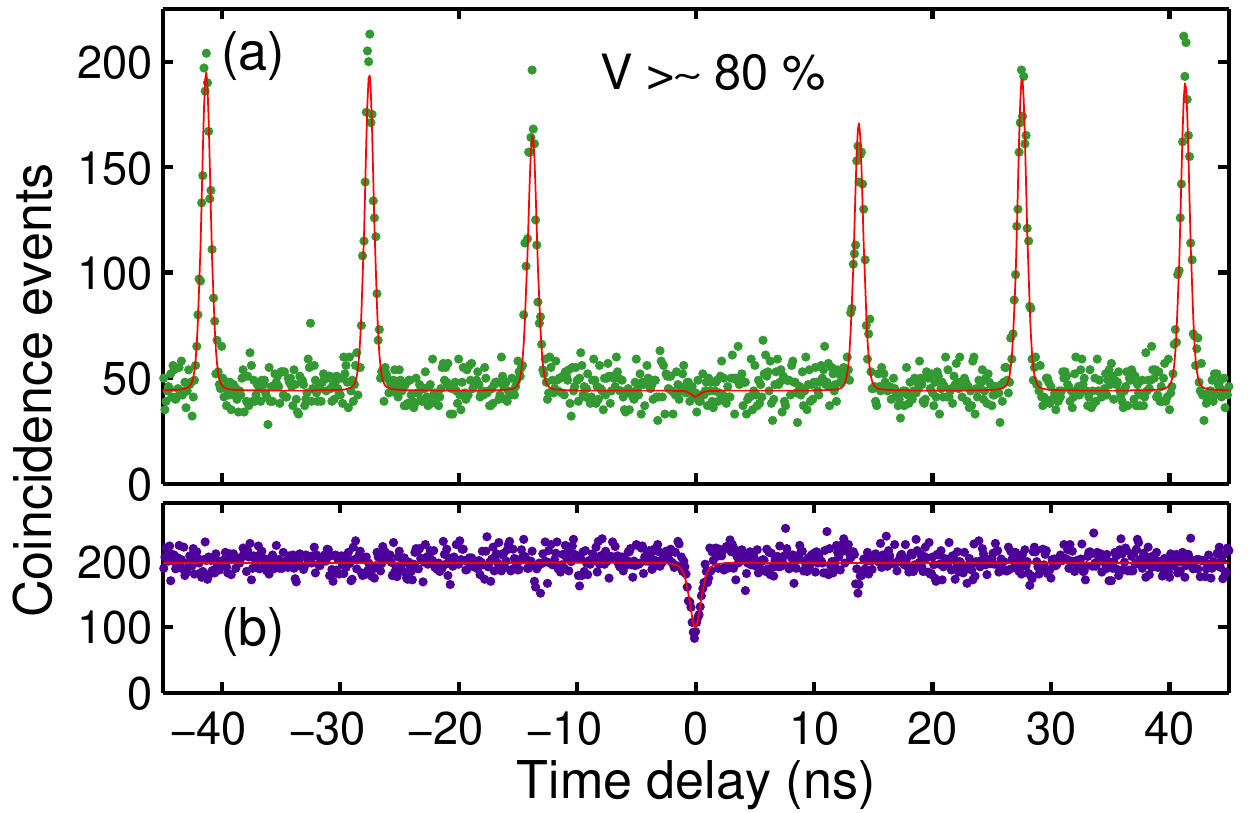}
\caption{HOM correlation histograms of Device~C in the case of strict resonant excitation with 100-\si{\pico\second}-long pulses at 0.8$P_{sat}$ (a) and with continuous-wave excitation at 0.5$P_{sat}$ (b). The operation temperature of the experiment \SI{1.58}{\kelvin}. The fit of the experimental data to theory is shown by the red line.
 }
\label{fig5}
\end{center}
\end{figure}

\begin{figure}
\begin{center}
\includegraphics[width=0.85\columnwidth]{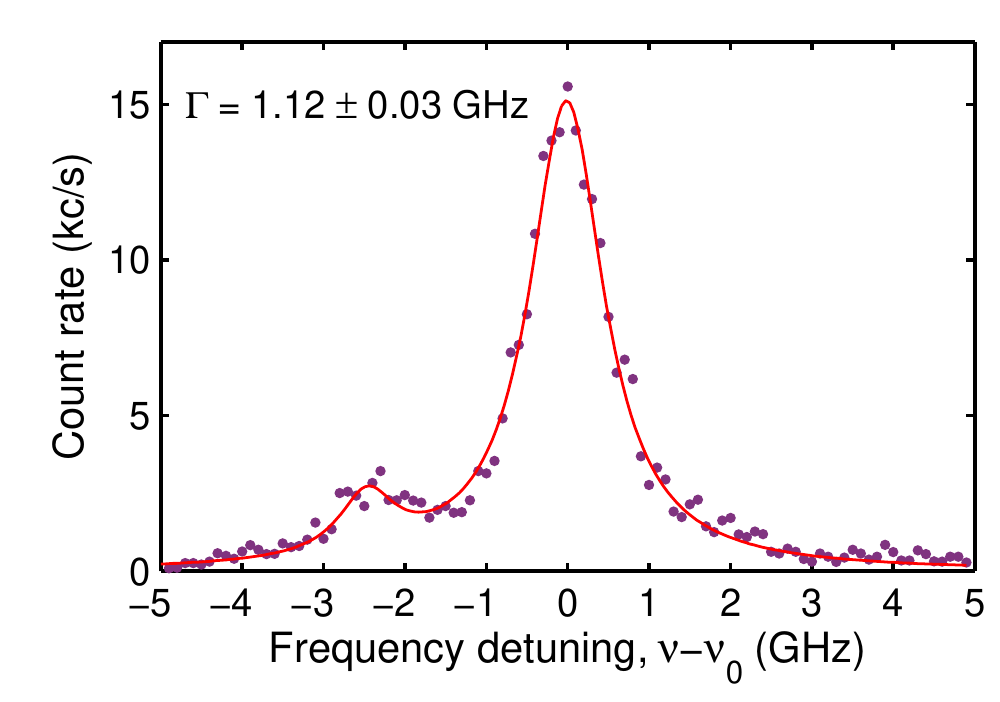}
\caption{Resonance fluorescence of a quantum dot versus laser frequency detuning. The data are recorded at a power of 0.2$P_{sat}$ at \SI{1.58}{\kelvin}. The resonance frequency $\nu = \SI{325.457}{\tera\hertz}$ corresponds to a wavelength of \SI{921.14}{\nano\meter}. The solid line is a Lorentzian fit to the data with linewidth $\Gamma = 1.12 \pm \SI{0.03}{\giga\hertz}$. The weak peak on the lower-energy side stems from another QD dipole weakly coupled to the waveguide.}
\label{figLinewidth}
\end{center}
\end{figure}

Strict resonant excitation may be implemented in order to achieve high indistinguishability without spectral filtering. Figure~\ref{fig5}(a) and (b) show the results of HOM measurements using 100-\si{\pico\second}-pulsed excitation and CW excitation, respectively. The high degree of indistinguishability is immediately visible in the pulsed data by the complete absence of a peak around zero time delay. However, in this case a quantitative extraction of a degree of indistinguishability from the pulsed data is complicated by two effects: the applied pulsed laser source had a CW background (at the level of \SI{23}{\percent}) and the relatively long excitation pulses imply that a double excitations of the QD occur with a non-negligible probability (leading to a contribution $g^{2}(0) \sim \SI{5.5}{\percent}$). The former effect leads to a finite background level in the data of Fig.~\ref{fig5}(a), which has the temporal dependence of the data in Fig.~\ref{fig5}(b). In the present analysis we attempt to include these background effects, cf.~Appendix B for the details of the analysis, and conclude that the degree of indistinguishability is $\gtrsim\SI{80}{\percent}$. We note that this is a conservative estimate since in the analysis it is assumed that the background originates solely from the CW fluorescence from the QD, while the finite extinction of the pump laser would provide a flat background and therefore lead to a higher extracted indistinguishability. From the resonant-excitation experiment, by scanning the laser frequency across the QD resonance and recording the integrated intensity on a CCD camera, we also extract a QD linewidth of $1.12 \: \mathrm{GHz}$ (cf.~Fig.~\ref{figLinewidth}), which is only 1.3 times wider than the natural linewidth. We emphasize that a transform-limited linewidth is a much stricter requirement than the demonstration of subsequently emitted indistinguishable photons, since the former is sensitive to slow charge noise~\cite{Kuhlmann2015}. This confirms the expectation that the extracted degree of indistinguishability of $\gtrsim\SI{80}{\percent}$ is a conservative lower bound.  Achieving a narrow linewidth is essential for fully exploiting the high cooperativity of the photon-QD interface \cite{Lodahl2015}.

\section{Summary}

In conclusion, we have experimentally demonstrated  high-purity and highly indistinguishable single-photon sources based on QDs embedded in planar nanoscale waveguides with integrated electrical contacts. The role of photon jitter in quasi-resonant excitation was identified, and a significant improvement in indistinguishability was observed when implementing spectral filtering or strict resonant excitation. A high indistinguishability of $\sim\SI{94}{\percent}$ was found, which illustrated the exciting potential of the planar waveguide platform. The indistinguishability can readily be further improved by enhancing the decay rate of the QD, which can be done by increasing the oscillator strength of the QD~\cite{Tighineanu2016a} or by implementing even modest Purcell enhancement~\cite{Grange2017}. Alternatively the temperature could be reduced further or the relevant decoherence phonon modes suppressed~\cite{Tighineanu2016}. A major issue for resonant excitation  schemes is to efficiently extinct the excitation laser from the signal, and here the planar platform may be advantageous since the pump beam could be guided laterally through the structure~\cite{Muller2007} or a vertical pump beam may be applied at one spot while collection is from another. This may overcome the intrinsic source efficiency limitations of \SI{25}{\percent} found in vertical devices when implementing cross-polarization extinction in excitation~\cite{Somaschi2016,Loredo2016,Wang2016,Ding2016}.  Combining these functionalities into a single device would eventually enable a fully deterministic and coherent single-photon source, which could subsequently be spatially demultiplexed by implementing fast switches to generate a scalable resource of single photons. The limit to the number of achievable simultaneous photons on demand with such an approach is ultimately determined by any loss processes that leads to an exponential reduction of the rate of photon generation. This constitutes an important future engineering challenge. Another important application area is for the development of deterministic spin-photon interfaces, where coherent light-matter interaction is a prerequisite for advanced quantum-network architectures~\cite{Gao2015}.

\begin{acknowledgments}
We acknowledge Kasper Prindal-Nielsen for the $\beta$-factor calculations, Camille Papon for the help with the characterization of the single-photon source efficiency, Liang Zhai for the $g^{2}(0)$ calculations, and Tim Schr{\"o}der for  valuable input to the experimental setup. We gratefully acknowledge financial support from the European Research Council (ERC  Advanced Grant ``SCALE''), Innovation Fund Denmark (Quantum Innovation Center ``Qubiz''), and the Danish Council for Independent Research. SIP \& JDS acknowledge support from the KIST flagship institutional program. AVK, IS, ML \& RJW acknowledge support  from SNF (project 200020\_156637) and NCCR QSIT. A.L. and A.D.W. gratefully acknowledge support of BMBF - Q.com-H 16KIS0109 and the DFG - TRR 160.
\end{acknowledgments}

\appendix
\section{Reliable extraction of photon indistinguishability from a pulsed Hong-Ou-Mandel experiment}

\begin{figure}
\begin{center}
\includegraphics[width=0.9\columnwidth]{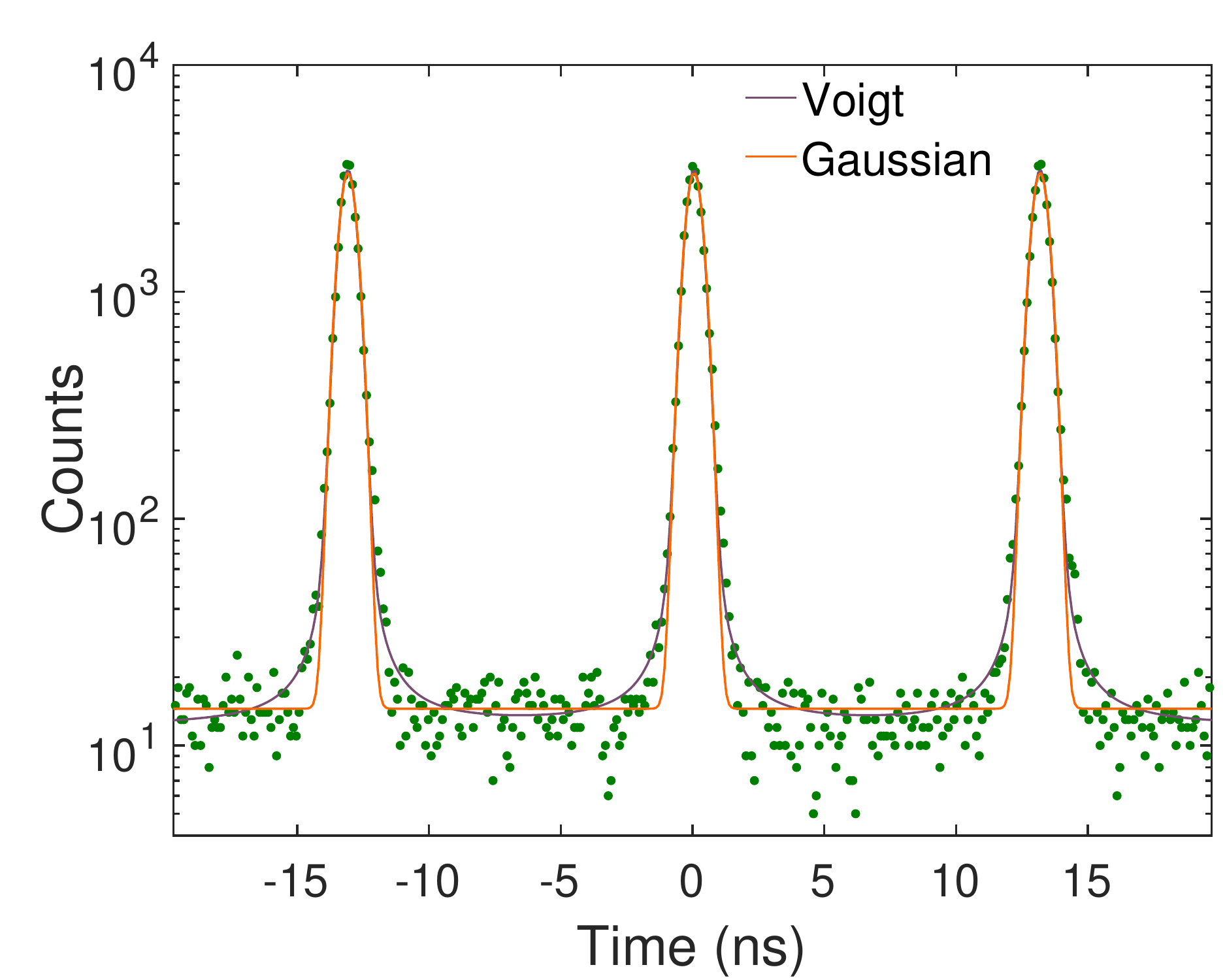}
\caption{The measured instrument response function (IRF) of the Hanbury Brown and Twiss setup fitted with both a Voigt (purple) and a Gaussian (orange) function using the Poission MLE described in the text. The Voigt function with its broader tail provide a superior fit to the line shape of the IRF evidenced by the smaller $\chi^2_{\mathrm{MLE}} =  2.3$ vs. $\chi^2_{\mathrm{MLE}} =  5.7$ for the Gaussian fit.}
\label{figIRFSM}
\end{center}
\end{figure}

In this section, we present the fitting routine used to model the Hong-Ou-Mandel data and demonstrate the key importance of applying appropriate noise statistics for the coincidence counts and the correct line shape for the peaks in order to reliably extract the photon indistinguishabily.

The pulsed two-photon correlation histograms in Fig.~\ref{fig4}(a) and (b) consist of a series of peaks. The peak amplitudes are determined by the relative probabilities for two photons to propagate along the different paths of a Mach-Zehnder interferometer, with either \SI{2.7}{\nano\second} or \SI{13}{\nano\second} delay lengths. The central peak near zero time delay $\Delta t = 0$ is the coincidence counts for two photons arriving on the two APDs after interfering on the last beam splitter. For completely indistinguishable photons the central amplitude is zero and the degree of indistinguishability can be quantified as the amplitude relative to the expected amplitude obtained for distinguishable photons.

As seen in the data in Fig.~\ref{fig4}(a) for a path difference of \SI{2.7}{\nano\second} the correlation histograms consist of a five-peak cluster repeated with a \SI{13}{\nano\second} period. The visibility in this case is given by \cite{Somaschi2016}
\begin{equation}
V_{2.7\mathrm{ns}}=\frac{R^2+T^2}{2RT}-\frac{2 A_0}{A_{{+2.7\mathrm{ns}}}+A_{-2.7\mathrm{ns}}},\label{eq:V27}
\end{equation}
where $A_0$ is the area of the central peak and $A_{\pm2.7\mathrm{ns}}$ is the area of the two neighbouring peaks with $R=0.46$ and $T=0.54$ being the reflectivity and transmission of the last beam splitter. We note that no correction for the small $g^2(0)<\num{0.006}$ is implemented in the analysis since it is an intrinsic property of the photon source and not the measurement apparatus. For the path difference of \SI{13}{\nano\second} the correlation histograms result in a set of peaks each separated by the inverse of the laser repetition rate of \SI{13}{\nano\second}. The visibility is here given as \cite{Loredo2016}
\begin{equation}
V_{13\mathrm{ns}}=\frac{R^2+T^2-A_0/A}{2RT}, \label{eq:V13}
\end{equation}
where $A$ is the average amplitude of peaks \SI{25}{\nano\second} or further away.

The five-peak cluster for \SI{2.7}{\nano\second} path difference consists of overlapping peaks resulting from the exponential QD decay with a finite rate that is on the order of the time separation between the excitation pulses. The coincidence counts therefore never reach zero, even for perfect indistinguishability, and the tail of the neighbouring peaks strongly influences the fitted amplitudes. Such data must therefore be modelled carefully in order to extract reliable values of the photon indistinguishability.

We implement a rigorous fitting routine, which takes into account the exponential decay of the emitter, the measured instrument response function (IRF), and Poissonian counting statistics. Each peak is modelled as a double-sided single exponential decay convoluted with the measured IRF. The repetition rate of the laser and the decay rate of the QD is obtained independently from the time-resolved measurements. Likewise, the ratio between pairs of neighbouring peaks is fixed by the measured beam splitter transmission and reflection. The remaining free parameters are the individual peak amplitudes, the Mach-Zehnder time delay, an overall time shift, and a background. The IRF is measured by sending a laser pulse through the  detection setup and the resulting peaks are fitted with a Voigt function and a background to obtain a background-free IRF used in the convolution. The Voigt function accounts for a longer tail present in the measured IRF that is not captured by a Gaussian fit (cf.~Fig.~\ref{figIRFSM}).

\begin{figure}
\begin{center}
\includegraphics[width=0.8\columnwidth]{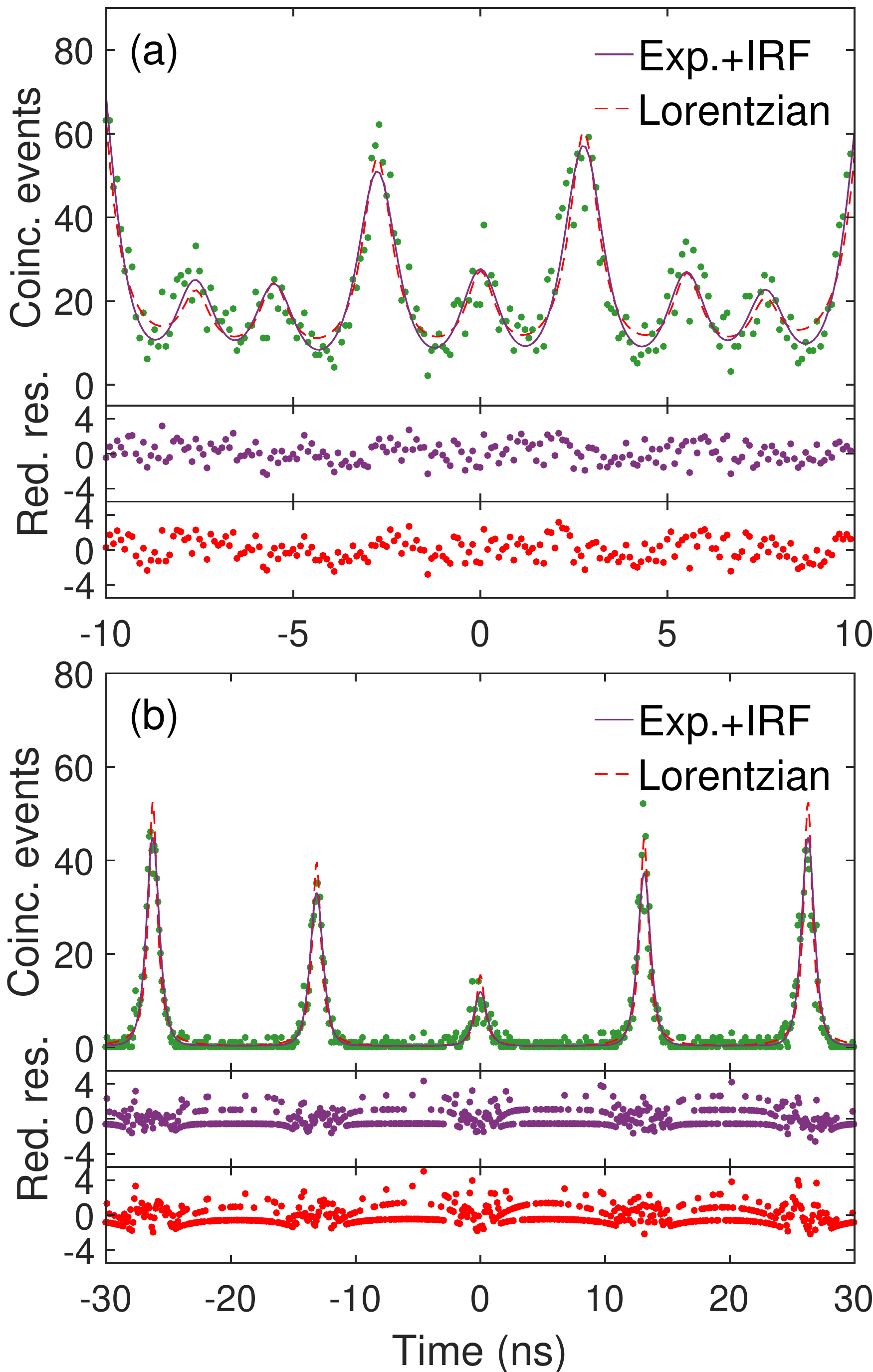}
\caption{Two-photon coincidence data at \SI{4.2}{\kelvin} for \SI{2.7}{\nano\second} (a) and \SI{13}{\nano\second} (b) path delay. The peaks are fitted with either double-sided exponentials convoluted with the measured instruments response (solid purple) or a Lorentzian (dashed red) by optimizing the Possion MLE. The corresponding reduced residues for each fit are shown below the figures. }
\label{fig3SM}
\end{center}
\end{figure}

We note that the photon indistinguishability data (cf.~Fig.~\ref{fig4}) only contain few coincidence counts in time bins in-between the peaks and more importantly, in the central peak, which contains the information about the indistinguishability. These low counts pose a challenge for traditional least square fitting routines that minimizes the reduced $\chi_\mathrm{red}^2$
\begin{equation}
 \chi_\mathrm{red}^2 = \frac{1}{n-\nu}\sum_{i=0}^n \frac{(y_i-f(\Delta t_i))^2}{y_i},\label{eq:chi2LS}
\end{equation}
where $n$ is the number of datapoints and $\nu$ is the number of free parameters. Here $y_i$ is the measured datapoint and $f_i\equiv f(\Delta t_i)$ is the fitting function associated with the data bin $\Delta t_i$. Minimizing Eq.~\eqref{eq:chi2LS} assumes a Gaussian noise distribution whereas the coincidence counts in reality follow a Poisson distribution \cite{Gregory2005}. Using traditional least square fitting procedures on such data is therefore known to lead to biases \cite{Laurence2010} where especially bins with $y_i=0$ cannot be properly handled as they cause $\chi^2$ to diverge.

\begin{table*}
\begin{tabular}{c@{\quad}SSSSSS}
\multicolumn{7}{c}{$\Delta t = \SI{2.7}{\nano\second}$ }\\
\toprule
 & \multicolumn{2}{ >{\columncolor{Gray}[2pt][-3pt]\centering}m{3.8cm} }{\textbf{Exp \& IRF/MLE}} & \multicolumn{2}{>{\centering}m{3.8cm}}{Lorz/LS} & \multicolumn{2}{>{\centering}m{3.8cm}}{Exp \& IRF/LS}\\
\cmidrule(r){2-3}
\cmidrule(r){4-5}
\cmidrule(){6-7}
$T$ (K) & {Visibility V}& {$\chi_\mathrm{mle}^2 $} & {Visibility V} & {$\chi_\mathrm{red}^2$} & {Visibility V} & {$\chi_\mathrm{red}^2$}\\
\midrule
\SI{4.2}{\kelvin}  & \num{0.62\pm 0.02} & 1.29 & \num{0.75\pm 0.04} & 1.82 & \num{0.65 \pm 0.03} & 1.42\\
\SI{10}{\kelvin}  & \num{0.54\pm 0.02} & 1.29 & \num{0.64\pm 0.05} & 1.82  & \num{0.56 \pm 0.04} & 1.46\\
\SI{15}{\kelvin}  & \num{0.40\pm 0.02} & 1.47 & \num{0.50\pm 0.05} & 1.83  & \num{0.43 \pm 0.04} & 1.62\\
\SI{20}{\kelvin}  & \num{0.20\pm 0.02} & 2.04 & \num{0.27\pm 0.05} & 2.00  & \num{0.23 \pm 0.04} & 2.28\\
\SI{25}{\kelvin}  & \num{0.21\pm 0.04} & 1.31 & \num{0.24\pm 0.07} & 1.67  & \num{0.22 \pm 0.06} & 1.53\\
\cmidrule(r){1-3}
\cmidrule(r){4-5}
\cmidrule(){6-7}
Methods in literature & \cite{Kambs2016} &  & \cite{Gazzano2013,Gschrey2015,Thoma2016} & & \cite{He2013,Madsen2014,Somaschi2016,Ding2016,Grange2017} & \\
\bottomrule
\end{tabular}
\\[1em]
\caption{Comparison of the fitting procedure used to analyse the data of the Hong-Ou-Mandel experiment when the peaks are fitted by either double-sided exponentials convoluted with the instrument response (Exp \& IRF/MLE) or Lorentzians (Lorz). The quoted errorbars are 95\% confidence intervals for the parameters extracted from the fits. The reported $\chi_\mathrm{mle}^2$ values are normalized by $1/(n-\nu)$. Literature references using the presented fitting routines to extract the HOM visibility in QDs. (Lorz---Lorentzian functions, LS---Least Square, MLE---Maximum Likelihood Estimator.)}
\label{tab:fit}
\end{table*}

To overcome the problem of low counts we instead use the maximum likelihood estimator (MLE) for the Poisson distribution and optimize that over the model parameters.
The probability of measuring $y_i$ coincidence events in the time bin $\Delta t_i$ when expecting $f_i$ on average is given by the Poisson distribution
\begin{equation}
 P(y_i|f_i) = \frac{f_i^{y_i}}{y_i!}\exp{-f_i}.\label{eq:likelihood}
\end{equation}
Assuming independent data points the global likelihood is
\begin{equation}
 \mathcal{L}(\mathbf{y}|\mathbf{f})= \prod_{i=1}^n P(y_i|f_i).
\end{equation}
To overcome underflow errors it is customary to minimize twice the negative logarithm of the normalized likelihood, i.e.
\begin{equation}
\begin{aligned}
\chi_\mathrm{mle}^2
	&=- 2 \ln\left(\frac{\mathcal{L}(\mathbf{y}|\mathbf{f})}{\mathcal{L}(\mathbf{y}|\mathbf{y})} \right) \\
	&= 2 \sum_{i=1}^n (f_i-y_i) - 2 \sum_{\substack{i=1\\y_i\neq 0}}^n x_i \ln\left(f_i/y_i\right).
\end{aligned}
\end{equation}
Similarly to the reduced $\chi_\mathrm{red}^2$ for least square fitting, the figure of merit $\chi_\mathrm{mle}^2$ approaches one after normalizing with the number of degrees of freedom $n-\nu$ when the model accurately describes the data. In the data analysis the global minimum optimization is run 50 times with a random set of initial parameters to ensure proper convergence since the algorithm is less robust than the Levenberg-Marquardt routine used for least square fitting.

Another important point is to choose the correct line shape of the correlation function to model the data. We compare in Table~\ref{tab:fit} the extracted values for the two-photon-interference visibility $V$ and the reduced $\chi_\mathrm{mle}^2$ when applying the correct exponential line shape and measurement IRF compared to the case of a heuristic Lorentzian function.  The "long-tail" of the Lorentzian function essentially imply that the extracted visibility $V$ is systematically and significantly overestimated in this case. The corresponding fits to the experimental data are reproduced in Fig.~\ref{fig3SM}(a).
Table~\ref{tab:fit} summarizes the outcome of the detailed analysis of our indistinguishability data with the different approaches applied in the literature.

For the \SI{13}{\nano\second} time delay we again compare our data with Lorentzian peaks as shown in Fig.~\ref{fig3SM}(b). For all the datasets the Lorentzian functions overestimate the peak heights and at the same time predict a too broad tail. As evident from the small reduced residues the data points above the fitted peaks for exponential fit are within the expected errorbars for Poissonian noise. The overestimated central peak leads to overall lower visibilities, despite the overestimated side peaks, and the broad tail causes the background to be pushed below zero. The latter is not physical since only positive values are allowed in the Poisson distribution.

\section{Analysis of the two-photon correlation histogram under resonant excitation with continuous-wave background}

The measured two-photon correlation histogram under strict resonant excitation is shown in Fig.~\ref{fig5}, and it is observed that the peak near zero time delay is virtually absent. However, an overall background arises from coincidence counts from photons emitted when the QD is excited by the residual CW laser. The background complicates the analysis of the indistinguishability measurements. Most importantly the CW HOM coincidence histogram exhibits a trough at zero time delay that masks the true peak height in the pulsed HOM histogram, which needs to be accounted for. This trough in the data cancels the residual peak in the pulsed indistinguishability measurement. For infinitely fast detector the trough would go to zero for a single QD, i.e. the depth and the width of the measured trough (Fig.~\ref{fig5}(b)) depends on the time response of the system, the lifetime, the coherence time, and $g^{2}(0)$ \cite{Proux2015}.

To estimate the peak amplitude at zero time delay $\tilde A_0$, we may subtract the trough depth, obtained from the CW HOM measurements. An estimate of the indistinguishability $\tilde V$ can then be obtained from Eq.~\eqref{eq:V13} by using the average amplitude of peaks at longer delay in combination with $\tilde A_0$. Hence, using the peak/trough area as done in the previous section, would significantly underestimate the indistinguishability, since the CW trough is wider than the peaks in the pulsed measurement. Instead we use the amplitudes of the peaks to extract $\tilde V$.

To estimate the trough depth $A_\mathrm{CW}$ the CW HOM data are modelled with a double-sided exponential trough convoluted with the measured instrument response function of the full system, $f_\mathrm{CW}(t) = A_\mathrm{CW} [1-\mathrm{exp}(-\gamma_\mathrm{CW}\abs{\Delta t})]\otimes \mathrm{IRF(\Delta t)}$ where the fitted rate is $\gamma_\mathrm{CW} = \SI{3.1\pm 0.2}{\per\nano\second} $. This functional form is an approximation to the real shape but is sufficient to extract the trough depth and it limits the number of free parameters.

The pulsed HOM data are modelled with a series of double-sided exponents convoluted with the instrument response, $f_\mathrm{P}(t)$, as described in Appendix~A, in addition to a background $A_\mathrm{bg}=\num{44.0}$. Since the data may have a trough from the CW contribution we allow for the central peak of the fit to have a negative amplitude. The decay rate is fitted to be $\gamma_\mathrm P = \SI{6.2\pm 0.2}{\per\nano\second}$, which is a free parameter since the lifetime of the QD is IRF limited. From these data we can use the peak amplitudes and backgrounds to extract the central peak and side peak amplitudes
\begin{equation}
\begin{aligned}
& \tilde A_0 = f_\mathrm{P}(0)-f_\mathrm{CW}(0)\frac{A_\mathrm{bg}}{A_\mathrm{CW}}= \num{18.9}, \\
& \tilde A = f_\mathrm{P}(2\Delta t)=\num{148.9},
\end{aligned}
\end{equation}
and from Eq.~\eqref{eq:V13} we get an estimate of the indistinguishability of $\tilde V \sim \num{75} \%$.

We emphasize that the above estimate is a lower boundary of the actual indistinguishability. Hence, the 100-\SI{}{\pico\second}-long excitation pulses imply that there is a finite probability to re-excite the QD since it may have emitted a photon before the excitation pulse is over. This effect leads to a non-zero $g^{2}(0)$ and increases the central peak height in the pulsed HOM measurement. This additional contribution to $g^{2}(0)$ is a consequence of the excitation scheme and not intrinsic to the source. We  account for it by including a correction for a  finite $g^{2}(0)$ in the expression for the two-photon interference visibility in Eq.~\eqref{eq:V13}. Following the procedure in Ref.~\onlinecite{Loredo2016} for expressing the expected peaks area and adding a correction factor $1+g^{2}(0)$ for the central peak area similarly to Ref.~\onlinecite{Santori2002}, we have
\begin{equation}
	 \tilde{A_0} / \tilde{A} = (R^2+T^2)(1+g^{2}(0))+2RTV,
\end{equation}
and we get a corrected expression for the two-photon interference visibility
\begin{equation}
 \tilde{V}_{13\mathrm{ns}} = \frac{(R^2+T^2)(1+g^{2}(0))-\tilde{A}_0/\tilde{A}}{2RT}.\label{eq:V13g2}
\end{equation}
The residual $g^{2}(0)$ due to re-excitation can be calculated numerically by including also the response of the applied etalon to spectrally filter the emitted photons. We find $g^{2}(0) \sim 0.055$. Using this value in  Eq.~\eqref{eq:V13g2}, we estimate that the indistinguishability of the photons obtained by resonant pulsed excitation is $\tilde V_{13\mathrm{ns}} \gtrsim \num{80} \%$.


%

\end{document}